%% file: msj2.tex
\documentclass{jkas}

\def\year{2018} 
\def\month{March} 
\def\volume{00} 
\def\issue{0} 
\def\beginpage{1} 
\def\endpage{99} 
\def\received{---} 
\def\accepted{---} 

\date{Received \received ; accepted \accepted}

\title{
KMT-2016-BLG-0212: First KMTNet-Only Discovery of a Substellar Companion
}

\author[1]{K.-H.~Hwang}
\author[1]{H.-W.~Kim}
\author[1]{D.-J.~Kim}
\author[1,2,3]{A.~Gould}
\author[4]{M. D. Albrow} 
\author[1,5]{S.-J. Chung} 
\author[6]{C. Han} 
\author[7]{Y. K. Jung} 
\author[1]{Y.-H. Ryu}
\author[7]{I.-G. Shin} 
\author[8,9]{Y.~Shvartzvald} 
\author[7]{J. C. Yee} 
\author[10,11]{W.~Zang} 
\author[12]{W. Zhu} 
\author[1,13]{S.-M. Cha} 
\author[1,5]{S.-L. Kim} 
\author[1,5]{C.-U. Lee} 
\author[1]{D.-J. Lee} 
\author[1,13]{Y. Lee} 
\author[1,5]{B.-G. Park} 
\author[3]{R. W. Pogge}

\affil[1]{Korea Astronomy and Space Science Institute, Daejon 34055, Republic of Korea}
\affil[2]{Max-Planck-Institute for Astronomy, K\"{o}nigstuhl 17, 69117 Heidelberg, Germany,
\email{gould@astronomy.ohio-state.edu }}
\affil[3]{Department of Astronomy Ohio State University,
140 W.\ 18th Ave., Columbus, OH 43210, USA}
\affil[4]{University of Canterbury, Department of Physics and
Astronomy, Private Bag 4800, Christchurch 8020, New Zealand}
\affil[5]{Korea University of Science and Technology, 
217 Gajeong-ro, Yuseong-gu, Daejeon 34113, Republic of Korea}
\affil[6]{Department of Physics, Chungbuk National University,
Cheongju 28644, Republic of Korea}
\affil[7]{Harvard-Smithsonian CfA, 60 Garden St.,Cambridge, MA 02138, USA}
\affil[8]{Jet Propulsion Laboratory, California Institute of
Technology, 4800 Oak Grove Drive, Pasadena, CA 91109, USA}
\affil[9]{NASA Postdoctoral Program Fellow}
\affil[10]{Physics Department and Tsinghua Centre for
Astrophysics, Tsinghua University, Beijing 100084, China}
\affil[11]{Department of Physics, Zhejiang University, Hangzhou,
310058, China}
\affil[12]{Canadian Institute for Theoretical Astrophysics, 
University of Toronto, 60 St George Street, Toronto, ON M5S 3H8, Canada}
\affil[13]{School of Space Research, Kyung Hee University,
Yongin, Kyeonggi 17104, Republic of Korea} 

\def\corrauthor{
Andrew Gould
}

\def\runningauthor{
Kyu-Ha Hwang, Hyoun-Woo Kim, et al.
}

\def\runningtitle{
First KMTNet-Only Discovery of a Substellar Companion
}

\def\keywords{
gravitational microlensing  -- planets 
}

\def\abstracttext{We present the analysis of KMT-2016-BLG-0212,
a low flux-variation $(I_{\rm flux-var}\sim 20$) microlensing event, which
is well-covered by high-cadence data from the three Korea Microlensing
Telescope Network (KMTNet) telescopes.  The event shows a short
anomaly that is incompletely covered due to the brief visibility intervals
that characterize the early microlensing season when the anomaly occurred.
We show that the data are consistent with two classes of solutions,
characterized respectively by low-mass brown-dwarf $(q=0.037)$ and sub-Neptune
$(q<10^{-4})$ companions, respectively.  Future high-resolution imaging should
easily distinguish between these solutions.
}

\def\muas{{\mu\rm as}}
\def\kpc{{\rm kpc}}

\def\eff{{\rm eff}}

\def\e{{\rm E}}

\def\apj{{ApJ}}
\def\aj{{AJ}}

\def\aap{{A\&A}}
\def\pasp{{PASP}}
\def\mnras{{MNRAS}}

\begin{document}
\jkashead 


\section{{Introduction}
\label{sec:intro}}

The Korea Microlensing Telescope Network (KMTNet, \citealt{kmtnet})
was originally designed to detect and characterize microlensing planets
without the need for followup observation.  

\citet{gouldloeb} had 
originally advocated a two-stage approach to finding microlens planets:
in the first stage, a low-cadence, wide-area survey operating from
a single site would detect 
microlensing events in real time and issue alerts to the microlensing
community, while in the second stage, a broadly distributed network
of narrow-angle telescopes would intensively monitor
individual events discovered in the first stage.  This strategy was
well-matched to the facilities that were available or were considered
feasible at that time.  Because microlensing events have typical
Einstein timescales $t_\e\sim 20\,$day, they can be reliably discovered
in surveys with cadences $\Gamma\sim 1\,{\rm day}^{-1}$.  However, because
the optical depth to microlensing $\tau\sim 10^{-6}$ is low (even in
the densest star fields toward the Galactic bulge), 10--100 square degrees
must be monitored to find a large number of events.  This is essential
for finding planets, because the probability of detecting a planet
within a given microlensing event is roughly $q^{1/2}$, where $q$
is the planet/host mass ratio.  For relatively common planets $q\sim 10^{-4}$
\citep{ob05169,gould10,ob07368,shvartzvald16,suzuki16,ob171434}, this 
probability is therefore 1\%.  That is, the probability that any given observed
star will give rise to a planetary signal is $\sim \tau\sqrt{q}\sim 10^{-8}$,
meaning that one must observe several $10^8$ stars to have a few events per
year with potential planetary signals.  On the other hand, to detect
a planetary signals requires a cadence $\Gamma$ that is sufficiently
high to characterize the brief planetary signal
$t_p\sim t_\e\sqrt{q}\rightarrow 5(q/10^{-4})^{1/2}\,{\rm hr}$.  
That is $\Gamma\sim 1\,{\rm hr}^{-1}$ would be required to characterize
``Neptunes'' and $\Gamma\sim 4\,{\rm hr}^{-1}$ would be required to detect
Earths \citep{henderson14}.

The strategy advocated by \citet{gouldloeb} was successful at finding planets,
beginning with OGLE-2005-BLG-071 \citep{ob05071}.  However, it was 
fundamentally limited by scarce telescope resources for ``stage two''
(followup) observations.  Indeed, it was only by focusing on high-magnification
events, as advocated by \citet{griest98}, that the method proved to be as
successful as it did.

Second generation surveys by the Microlensing Observations in Astrophysics
(MOA, 2006+) and the Optical Gravitational Lensing Experiment (OGLE, 2010+) 
teams were able to cover very large areas at high cadence 
$\Gamma = 1$--$4\,{\rm hr}^{-1}$, and therefore became capable of
both finding microlensing events and characterizing planets without the
need for followup observations.  For example, \citet{ob120406}
were able to find and characterize OGLE-2012-BLG-0406Lb based on OGLE data
alone.  Moreover, by combining the OGLE, MOA, and Wise surveys,
\citet{shvartzvald16} were able to conduct a survey-only microlensing-planet
search with 24-hour coverage, albeit over a limited area.

By combining three 1.6m telescopes with $4\,\deg^2$ fields of view on
three continents (CTIO, Chile (KMTC), SAAO, South Africa (KMTS), and
SSO, Australia (KMTA)), KMTNet is able to monitor about 
$12\,\deg^2$ at $\Gamma=4\,{\rm hr}^{-1}$,
$41\,\deg^2$ at $\Gamma>1\,{\rm hr}^{-1}$, 
$85\,\deg^2$ at $\Gamma>0.4\,{\rm hr}^{-1}$, and
$97\,\deg^2$ at $\Gamma>0.2\,{\rm hr}^{-1}$, 
making it sensitive to, respectively, Earth-mass, Neptune-mass, Saturn-mass, 
and Jupiter-mass planets over these areas.  In the past, this capability
has led to the discovery of planets whose perturbations were either 
inadequately covered (e.g., \citealt{ob160613}) or not covered at all
(e.g., \citealt{ob170173}) by other surveys.  However, in these other
cases, the event itself was discovered by other surveys, and the planet
was found by examining KMTNet data.

Although such combined discoveries are an important contribution,
KMTNet also has the potential to independently discover microlensing
events, including some with planets and other interesting companions.
To date, KMTNet has focused on finding completed events \citep{2015event}
rather than issuing real-time alerts of ongoing events, as for example
OGLE has been doing for almost 25 years \citep{ews1}.  As discussed by
\citet{2015event}, this partly reflects the original design of KMTNet
as narrowly focused on four fields, for which it had been expected that
there would be relatively little role for real-time followup observations.
However, even taking account of the revised strategy described above,
the decision was still made to focus initially on completed events,
simply to catch up with the rapidly accumulating KMTNet light-curve 
database.  While this approach is finding many events
that were previously identified by other groups (and so, usually,
already examined in KMTNet data), it is also yielding a substantial
number of new events that were missed by OGLE and MOA for a variety
of reasons, usually non-coverage or coverage that was not adequate
to reliably identify the event.

\citet{2015event} developed a new algorithm for finding completed
events and applied it to the 2015, i.e., commissioning-year, KMTNet data.  
In 2016, the observation strategy was substantially changed to that above,
covering a roughly six times larger area than the 2015 survey.  
In addition, the algorithm
was upgraded to enable event detection from combined
light curves of all three observatories and of overlapping fields.
Both changes are likely to lead to a substantial increase in the number
of ``new events'' detected by KMTNet.  However, they also both contributed
to considerable delay in the public release of the 2016 data relative
to the KMTNet goal of releasing within six months of the end of each
season \citep{2015event}.  Hence, for 2016, we focused initially on
an expedited release \citep{2016event} of KMTNet events in the 
{\it Kepler}-{\it K2} Campaign 9 field \citep{gouldhorne,henderson16}.
This is not a particularly promising domain for new KMTNet events simply
because it was heavily covered by all microlensing surveys, including several
that were created especially to support {\it Kepler}-{\it K2} C9.
Nevertheless \citet{2016event} do report a number of such discoveries,
including KMT-2016-BLG-0212.

\section{{Observations and Event Recognition}
\label{sec:obs}}

KMT-2016-BLG-0212 lies at equatorial coordinates 
(RA, Dec)$=$(17:53:45.42,$-29$:05:12.80), corresponding to Galactic
coordinates $(l,b)=(0.79,-1.60)$.  It therefore lies in two overlapping
KMTNet fields, BLG02 and BLG42.  Because these fields strongly overlap
the {\it K2} C9 footprint, they were observed at a higher-than-usual
combined cadence $\Gamma=6\,{\rm hr}^{-1}$ from KMTS and KMTA beginning 
April 23 (${\rm HJD}^\prime \equiv {\rm HJD}-2450000 \sim 7501$).  However,
because the event had already essentially returned to baseline by this
date, this enhanced cadence has almost no practical importance for the
present study.  Hence,
the relevant observations were basically all taken at the standard
cadence for BLG02/BLG42 in 2016, $\Gamma=4\,{\rm hr}^{-1}$.  As mentioned
in Section~\ref{sec:intro}, KMTNet observations are carried out with
three identical 1.6m telescopes, each equipped with a $4\,{\rm deg}^2$
camera.  The cameras are each comprised of four chips (K,M,T,N).  The
event is located near the chip boundaries so that, by chance, it falls
in BLG02T, but in BLG42K, within these two slightly offset fields.
The great majority of data were taken in the $I$ band, with about
10\% of the KMTC images and 5\% of the KMTS images taken in the $V$ band,
solely to determine the colors of microlensed sources.  For the
light curve analysis, the data were reduced using the pySIS software
package \citep{albrow09}.

KMT-2016-BLG-0212 was originally recognized as ``possible microlensing''
in the summer of 2017, during a human review of $\sim 5\times 10^5$ candidates
found by an automatic classifier from among $\sim 3\times 10^8$ light curves.
The light curves were obtained from difference image analysis (DIA)
as implemented using publicly available code from \citet{wozniak}.
Whenever possible, the DIA input catalog is extracted from the OGLE-III
star catalog \citep{oiiicat}.  Otherwise, much shallower, 
DoPhot \citep{dophot} catalogs are derived from KMTNet images.
KMT-2016-BLG-0212 was identified on the light curve of an $I=19.2$
OGLE-III catalog star.  It was judged as ``possible'' rather than ``clear''
microlensing because its amplitude is relatively low and the light curve
is relatively noisy.  Indeed for these reasons, the algorithm found
$\Delta\chi^2=1521$ (relative to a flat line), which is fairly close
to the $\Delta\chi^2=1000$ threshold.  The anomaly that we will
investigate in this paper is not discernible in the DIA light curve, for
reasons that we discuss immediately below.

The anomaly was discovered in the course of routine vetting for false 
positives of all candidates that had been identified in the human review,
which was the final step in preparation for the KMTNet-{\it K2} data release
\citep{2016event}.  This review consisted of viewing side-by-side, 
an automated pySIS re-reduction of the light curve in the neighborhood
of the putative event and a 2016-2017 joint DIA light curve.

The anomaly is quite obvious in the pySIS reductions, which have substantially
less noise than the DIA reductions.  This is partly because the pySIS
``kernel'' is better matched to the point spread function (PSF), but mainly
because the input catalog star is displaced from the true position of
the microlensed source by $0.45^{\prime\prime}$.  In particular, the KMTA data that
contain the anomaly had exceptionally good seeing (for SSO), as low
as FWHM$\sim 1.3^{\prime\prime}$, which (due to the astrometric offset)
led to a poor fit, as recorded by the program, and so to photometry
that was even noisier than usual.  In brief, the re-reductions were
essential to the discovery of the anomaly.  Note that although the
automated pySIS reductions are quite good, the final pySIS reductions
were carried out by hand for optimal photometry.

\section{{Light Curve Analysis}
\label{sec:anal}}

\subsection{{Heuristic Analysis}
\label{sec:heuristic}}

\begin{figure}[h]
\centering
\includegraphics[width=90mm]{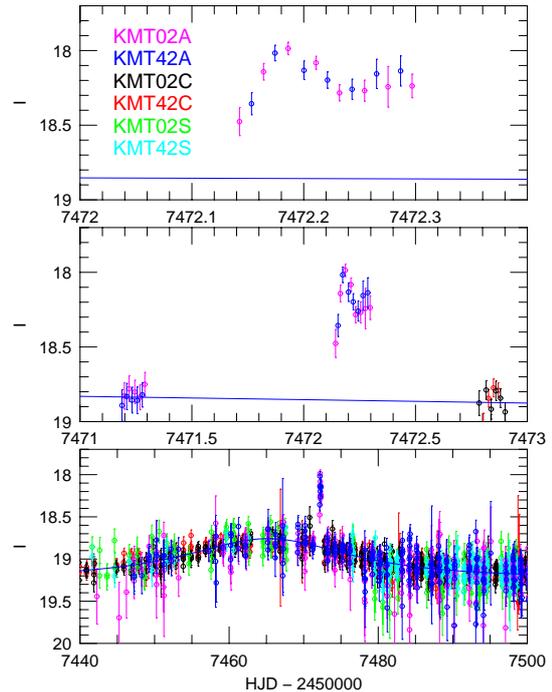}
\caption{Lightcurve and with single lens (1L1S) model for KMTNet observations
of KMT-2016-BLG-0212.  The top panel shows the caustic crossing, which is 
excluded from the fit, while the middle panel shows surrounding regions.
The magnitude of the flux variation, 
$I_{\rm flux-var} \equiv -2.5\log(10^{-0.4 I_{\rm peak}}-10^{-0.4 I_{\rm base}})\sim 20$
is quite low by the standards of published microlensing events.  Here
$I_{\rm peak}=18.8$ and $I_{\rm base}=19.2$ are resepectively the peak and baseline
of the underlying \citet{pac86} event.
\label{fig:lc}}
\end{figure}[h]

Figure~\ref{fig:lc} shows the KMTNet data with a single-lens single-source
(1L1S) model, for which the caustic-region (top panel) data are excluded.
The overall characteristics of these data are broadly similar to
those of OGLE-2017-BLG-0373 \citep{ob170373}: a low-amplitude event
with a short, incompletely covered anomaly that appears to be
consistent with a planetary caustic.  For that event, \citet{ob170373}
found that there were five different topologies that were roughly
consistent with the data, although in the end all but one of these were 
excluded at $\Delta\chi^2\gtrsim 100$.  In the present case, the interior
of the caustic appears to be more completely covered, but in contrast
to OGLE-2017-BLG-0373, neither the caustic entrance nor exit is fully
covered.  Thus, we proceed cautiously to evaluate all potentially viable
topologies.  As in the case of OGLE-2017-BLG-0373, we begin with a
heuristic analysis of the event \citep{gouldloeb}.

The point-lens fit yields \citet{pac86} parameters
$(t_0,t_\eff,t_\e)=(7465.2,11.2,17.1)\pm(0.2,1.2,2.9)\,$day.
Here, $t_0$ is the time of lens-source closest approach, $t_\e$ is the
Einstein crossing time, $t_\eff \equiv u_0 t_\e$
is the effective timescale, and $u_0$ is the
impact parameter (normalized to the angular Einstein radius $\theta_\e$).
 From Figure~\ref{fig:lc}, the perturbation is centered at 
$t_{\rm anom} \simeq {\rm  HJD}^\prime = 7472.3$, implying an offset from the
peak of $\delta t= + 7.1\,$day.  Therefore, if this perturbation is
due to a planetary anomaly, then the angle of the source
trajectory relative to the binary axis is 
$\alpha = \tan^{-1}(t_\eff/\delta t) = 57.6^\circ \pm 3.0^\circ $, and
the lens-source separation at the time of the anomaly is
$u_{\rm caust} = \sqrt{t_\eff^2+(\delta t)^2}/t_\e = 0.78\pm 0.16$.
We can then evaluate $s$, the projected separation of the host and companion
normalized to $\theta_\e$, from $|s-s^{-1}|= u_{\rm caust}$, which yields either
$s = 1.46\pm 0.11$ or $s = 0.68\pm 0.05$.  Naively, the anomaly in
Figure~\ref{fig:lc} ``looks like'' a major-image planetary perturbation.
Then following the analysis of \citet{ob170373} of OGLE-2017-BLG-0373,
we note that the above value of $\alpha$ would imply a diagonal
caustic crossing and hence a caustic-crossing size best-estimated from
the minor diameter $\Delta\eta_c = \sqrt{16 q/(s^4+s^2)} = 1.55 q^{1/2}$
\citep{han06}.  The caustic coverage is incomplete, but appears to be
slightly more than half over when the KMTA data end.  We therefore estimate
$t_{\rm caust}=0.3$ days, from which we derive 
$q=((t_{\rm caust}/t_\e)/1.55)^2=1.3\times 10^{-4}$.

\subsection{{Grid Search}
\label{sec:grid}}

The exercise in Section~\ref{sec:heuristic} shows, based on cursory
inspection of the light curve, that there is likely to be a $q\sim 10^{-4}$
major-image solution, but it does not show that this solution is either
unique or best.  Indeed, \citet{ob170373} showed that for the qualitatively
similar case OGLE-2017-BLG-0373, there were four additional topologies
that yielded viable fits to the data.  

\begin{figure}[h]
\centering
\includegraphics[width=80mm]{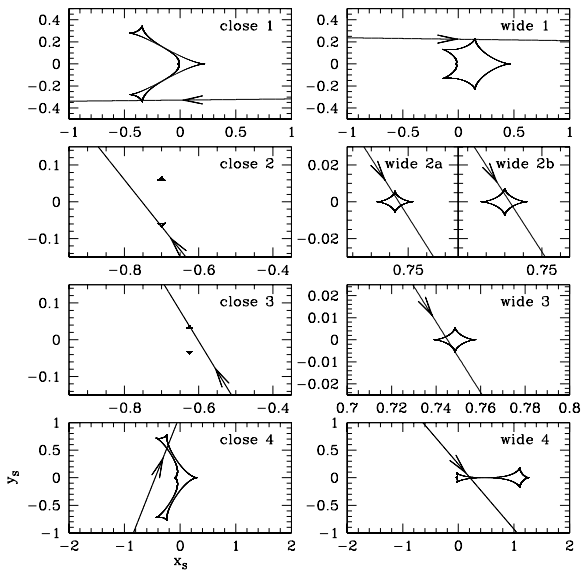}
\caption{Source trajectory and caustic geometries for nine solutions,
representing seven different topologies.
\label{fig:caust}}
\end{figure}

We therefore undertake a systematic grid search to find all such topologies.  
We first hold $(s,q)$ fixed at $100^2$ pairs of values
[$(-1\leq\log s \leq 1)\times(-5\leq\log q\leq 0)$], while seeding
the other parameters at $(t_0,u_0,t_\e)$ as derived above, $\rho=10^{-3}$,
and $\alpha$ at 10 equally spaced values around a circle.  We employ
Markov Chain Monte Carlo (MCMC) $\chi^2$ minimization to find the best
grid-point model.  We then seed new MCMCs with local minima on the $(s,q)$
plane derived from this grid search.  We find that there are six
other viable topologies (in addition to the one heuristically derived
in Section~\ref{sec:heuristic}).  Moreover, very similar to 
OGLE-2017-BLG-0373, we find two different geometries (``wide 2'' and ``wide 3'')
within the topology
identified in Section~\ref{sec:heuristic}).  We further divide ``wide 2''
into ``wide 2a'' and ``wide 2b'' because this broad minimum in the $\chi^2$
surface weakly separates into two sub-minima.
Figure~\ref{fig:caust}
shows the source trajectories for these nine different solutions.
[h]

\subsection{{Elimination of Some Topologies}
\label{sec:elim}}

\input tab_c

\input tab_w

\input tab_quant

\begin{figure}[h]
\centering
\includegraphics[width=80mm]{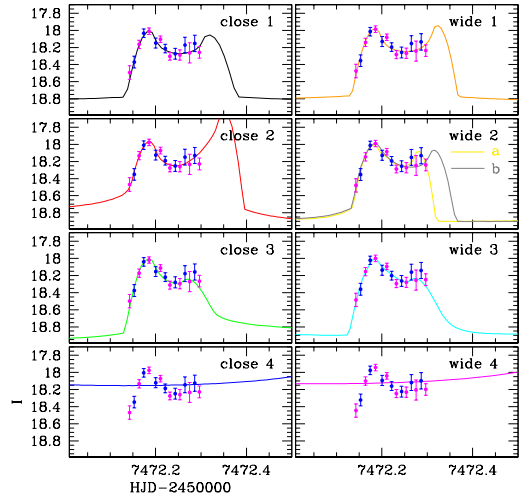}
\caption{Zoom of fits for nine different model geometries of
KMT-2016-BLG-0212 over the anomaly.  Solutions ``close 4'' and ``wide 4''
have poor fits and are excluded.
\label{fig:zoom}}
\end{figure}

These nine solutions are given in 
Tables~\ref{tab:close} and \ref{tab:wide}.  Three of these solutions
(``close 2'', ``close 4'' and ``wide 4'') 
have $\chi^2$ values that are substantially higher than the others.
Figure~\ref{fig:zoom}, which shows the light-curve fits over the anomaly,
implies that a major reason for this is a very poor fit of the
latter two (``close 4'' and ``wide 4'') to the anomaly.
We consider that these are eliminated.
The remaining solutions fit the anomaly reasonably well.
[h]

\begin{figure}[h]
\centering
\includegraphics[width=80mm]{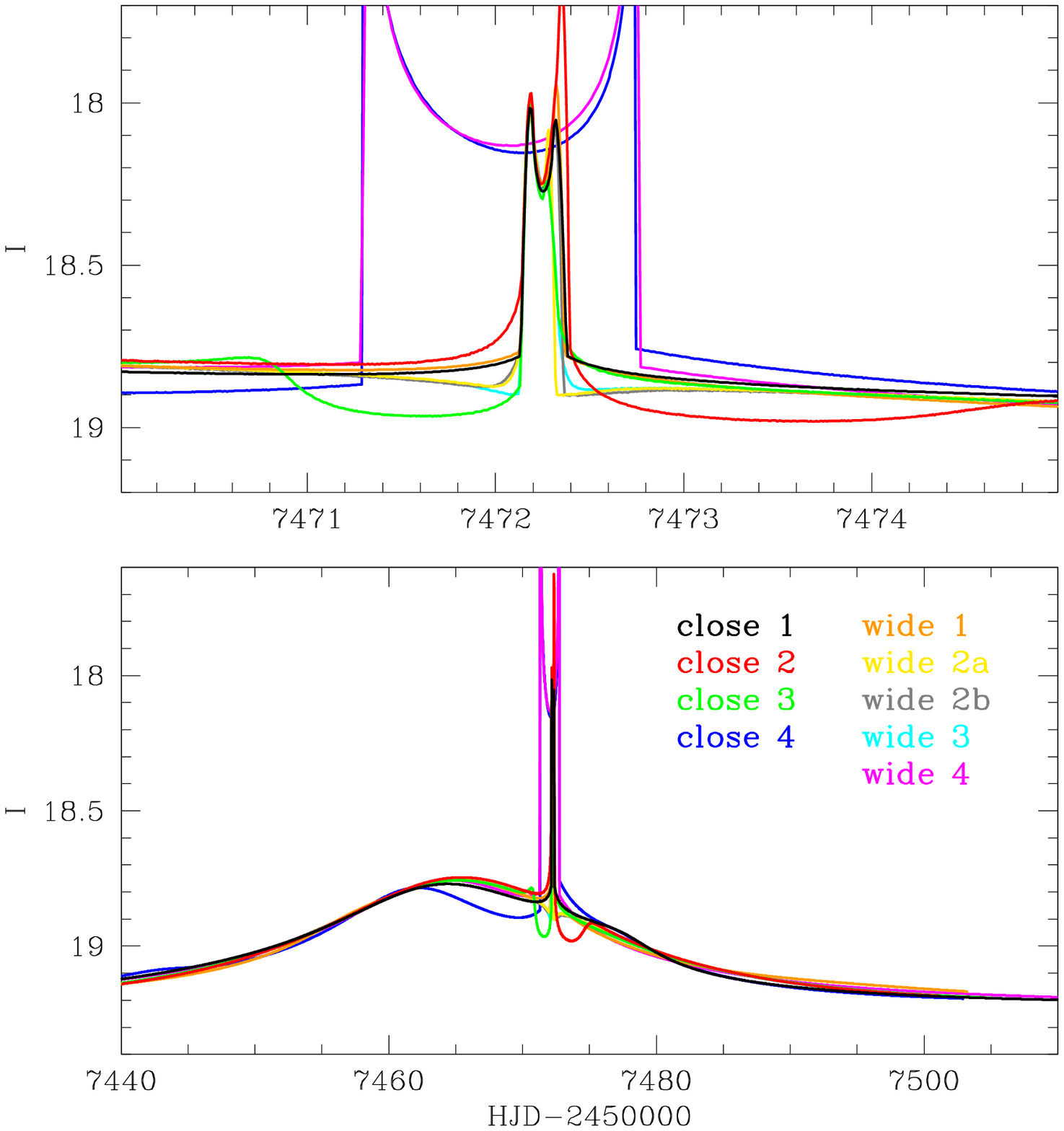}
\caption{Comparison of nine models (without data) that are broadly
compatible with the data.  The upper panel is a zoom of the region
near the caustic.
\label{fig:models}}
\end{figure}

\begin{figure}[h]
\centering
\includegraphics[width=80mm]{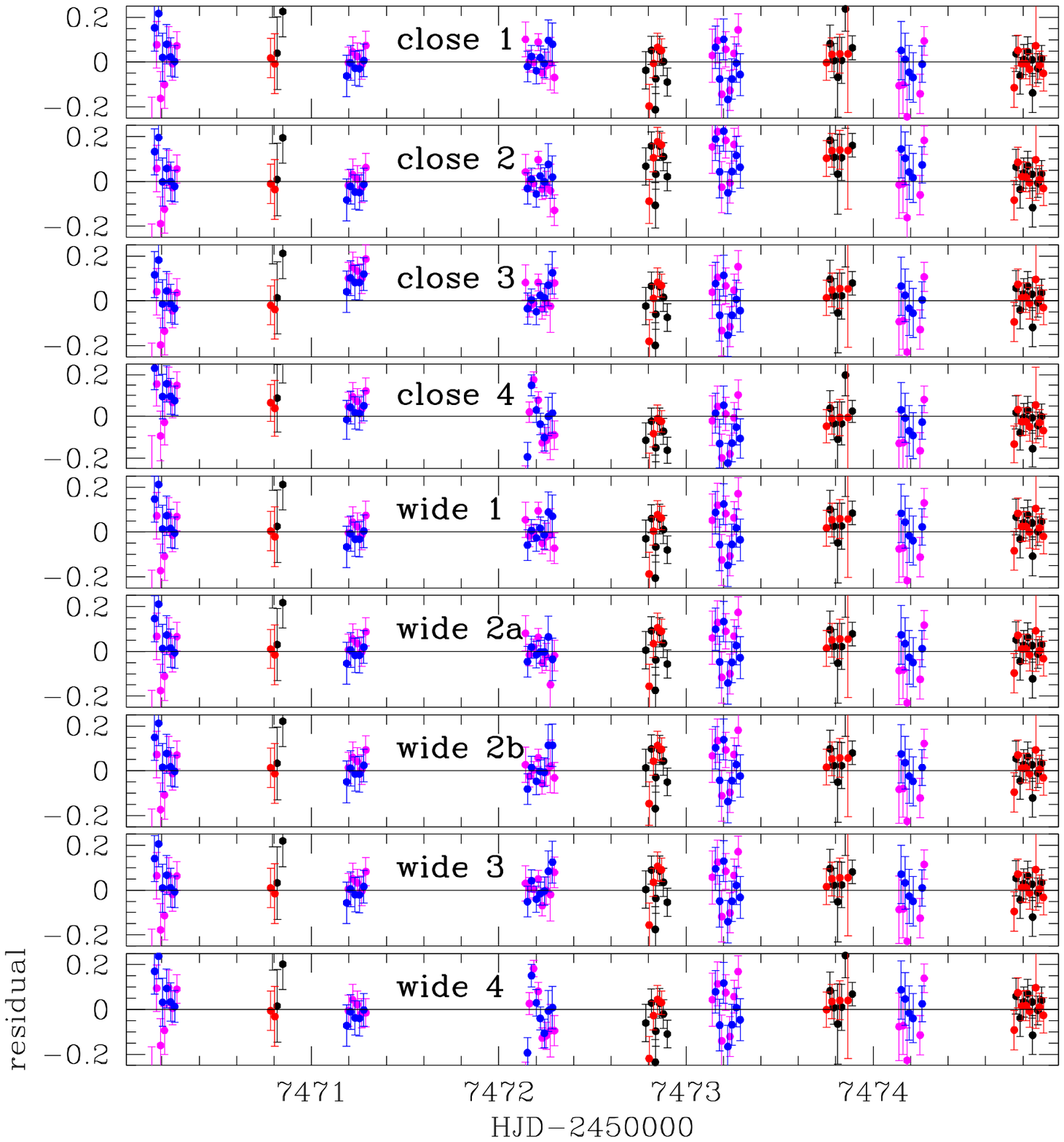}
\caption{Residuals of the data relative to the nine models shown in
Figure~\ref{fig:models}.  The quite poor match of model ``close 2''
during the 1.5 days centered on HJD$^\prime\sim 7473.5$ explains
the high $\chi^2$ of this model.
The origin of the relatively high $\chi^2$
of ``close 3'' model is the systematic deviation of the model in KMTA
data near HJD$^\prime\sim 7471.2$.  The mismatch of models ``close 4''
and ``wide 4'' are noticeable here but are more apparent in 
Figure~\ref{fig:zoom}.
\label{fig:residuals}}
\end{figure}

Figure~\ref{fig:models} shows the overall form of the nine models, and
Figure~\ref{fig:residuals} shows the residuals of the data for each model.
[h]
[h]
Figure~\ref{fig:residuals} shows that the high $\chi^2$ of model ``close 2''
is due to systematically high residuals during four consecutive episodes
of KMTC, KMTA, KMTC, KMTA observations beginning HJD$^\prime\sim 7472.8$,
which is explained by the long post-caustic ``dip'' of this model
in Figure~\ref{fig:models}.  It also shows that the
relatively high $\chi^2$ of
model ``close 3'' is primarily due to systematic residuals near 
HJD$^\prime\sim 7471.2$.
Comparing to Figure~\ref{fig:models}, we see that this is due to the
strong ``dip'' in this model just prior to the caustic crossing.
Finally, we note that although ``wide 1'' has even higher $\chi^2$ than
``close 3'', there are no strong residuals within the range displayed
in Figure~\ref{fig:residuals}.  The main problem for this model comes
from its long ``relative trough'' (compared to ``close 1'') after the
caustic exit, $7473\lesssim {\rm HJD^\prime}\lesssim 7480$.  
See Figure~\ref{fig:models}.  This issue also impacts ``close 3'',
albeit at a lower level.

\subsection{{Summary of Surviving Models}
\label{sec:survive}}

This series of rejections leaves models ``close 1'', 
``wide 2a'', 
``wide 2b'', and 
``wide 3'',
which have mass ratios, 
$q=3.7\times 10^{-2}$,
$q=4.9\times 10^{-5}$,
$q=8.3\times 10^{-5}$,
and
$q=4.8\times 10^{-5}$,
respectively.
The first solution (``Class I'')
which, depending on the host mass, could be a brown dwarf
or a high-mass planet, is preferred over the other three by 
$\Delta\chi^2\geq 6.8$.
Hence, it is favored, but not decisively.  The other three solutions
have $q\lesssim 10^{-4}$.

This second class of solutions (``Class II'')
are part of the same topology, namely the
one that was naively investigated in Section~\ref{sec:heuristic}.
Comparison to Table~\ref{tab:wide} 
shows that the simple reasoning in that section
predicted the parameters of these solutions reasonably well.  

This event is similar to the case of OGLE-2017-BLG-0373 \citep{ob170373}.
Also similar to that case, there are multiple geometries within this
topology that are qualitatively similar but can differ significantly
in the mass ratio $q$.  However, what is fundamentally different
about the present case is that one of the alternate topologies (which were
not anticipated by the naive reasoning of Section~\ref{sec:heuristic})
is competitive with (and indeed slightly preferred over) the naive
solution.

We note, however, that the two classes of solutions differ by a factor
2.5 in their source flux $f_s$, i.e., by $\sim 1\,$mag in
source magnitude (see Section~\ref{sec:cmd}).  
As we discuss in Section~\ref{sec:resolve}, this
will ultimately enable one to distinguish between these two classes
of solutions.

\section{{Physical Parameters}
\label{sec:phys}}

As just discussed, there are two classes of solutions with very different
topologies and very different planet-host mass ratios $q$.  The first
class has only one local minimum (``close 1''), with $q=0.037$.
The second class has three local minima (``wide 2a'', ``wide 2b'', ``wide 3''),
with $q$ ranging from $4.8\times 10^{-5}$ to $8.3\times 10^{-5}$.
For the second class, all the remaining parameters are essentially the
same with the exception of $\rho$, and even the three values of $\rho$ 
are basically consistent with one another within their rather large
errors.  See Table~\ref{tab:wide}.  Therefore, there are likewise
two classes of physical parameters for the host, with a factor
$\sim 1.8$ range in planet-host mass ratio within the second class.

\subsection{{Color-Magnitude Diagram}
\label{sec:cmd}}

\begin{figure}[h]
\centering
\includegraphics[width=90mm]{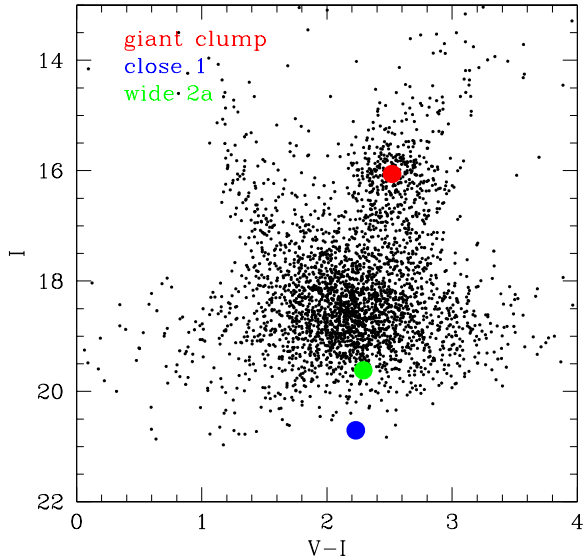}
\caption{Color-magnitude diagram (CMD) from pyDIA reductions of
KMTNet data, calibrated to OGLE-III \citep{oiiicat}.  The positions
of the clump and of the source for two of the solutions 
(``close 1'' and ``wide 2a'') are shown.  The source positions for the other
two ``Class II'' solutions (``wide 2b'' and ``wide 3'') are nearly
identical to ``wide 2a''.
\label{fig:cmd}}
\end{figure}

The first step toward estimating the physical parameters is to locate
the source star on a color-magnitude diagram.  The source color
should be independent of the model and should, in fact, be measurable
without reference to any model, i.e., by regression.  However, this
proves not to be the case for KMT-2016-BLG-0212.
In 2016, KMTNet took $V$-band data from
KMTC and KMTS.  Since the source lies in two overlapping fields
(BLG02 and BLG42), the source color can in principle be determined
independently from four different data sets.  However,
the faintness of the source in $V$-band
and the low-amplitude of the event together render regression-based
$(V-I)$ color estimates unstable.  
Hence, we must measure both the source color and magnitude
from each of the four $V/I$ data sets within the framework
of specific models.  We perform a special set of pyDIA reductions of the data
(i.e., different from the pySIS reductions from the main light-curve
analysis) because these simultaneously yield field-star photometry
on the same system as the light curve.  Unfortunately, the $V$-band
light curve from KMTS02 is not usable.  Hence, for each of the four
surviving models,  we have three independent measures of the source color
$(V-I)_s$ and four independent measures of the source magnitude $I_s$.
In each case, we find the offset of these quantities from the clump.
In Table~\ref{tab:quant} we present the means and standard errors of the mean
for these three (color) or four (magnitude) measures.  Figure~\ref{fig:cmd}
shows the pyDIA color-magnitude calibrated to OGLE-III \citep{oiiicat}
including the position of the red clump and the locations of the source
for the two classes of solution.

[h]

We see from these results that from the standpoint of the source
color and magnitude, there are essentially two classes of solutions,
close BD-class companion (Class I) and wide sub-Neptune-class companion 
(Class II).
We adopt the dereddened clump color and magnitude 
$[(V-I),I]_{\rm clump,0} = (1.06,14.44)$ from \citet{bensby13} and \citet{nataf13},
convert from $V/I$ to $V/K$ using the color-color relations of \citet{bb88},
and then apply the color/surface-brightness relations of \citet{kervella04}
to obtain 
\begin{equation}
\theta_* = 0.51\pm 0.09\,\muas\quad ({\rm I}),
\quad
\theta_* = 0.89\pm 0.19\,\muas\quad ({\rm II})
\label{eqn:thetastar}
\end{equation}
Then using the values (and errors) of $\rho$ and $t_\e$ from 
Tables~\ref{tab:close} and \ref{tab:wide}, one obtains
the Einstein radius $\theta_\e=\theta_*/\rho$ and 
proper motion $\mu=\theta_\e/t_\e$, as given in Table~\ref{tab:quant}.
We note that these two physical quantities are unusually
poorly constrained.  This is partly due to the large errors
in $\theta_*$, which is caused by the relatively poor measurement of $(V-I)_s$,
and partly due to the large errors in $\rho$, which is caused by
the incomplete coverage of the caustic entrance.

\begin{figure}[h]
\centering
\includegraphics[width=90mm]{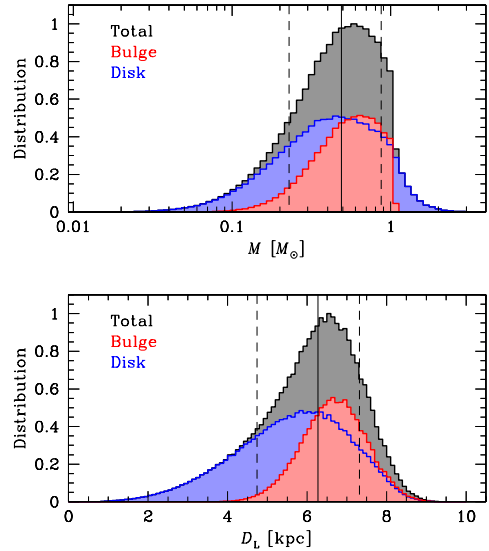}
\caption{Bayesian estimates, based on the \citet{han95} Galactic model
for the host mass and system distance of KMT-2016-BLG-0212 for
the ``close 1'' (BD-class) solution.  The distributions are
quite broad.
\label{fig:bayesc1}}
\end{figure}

\begin{figure}[h]
\centering
\includegraphics[width=90mm]{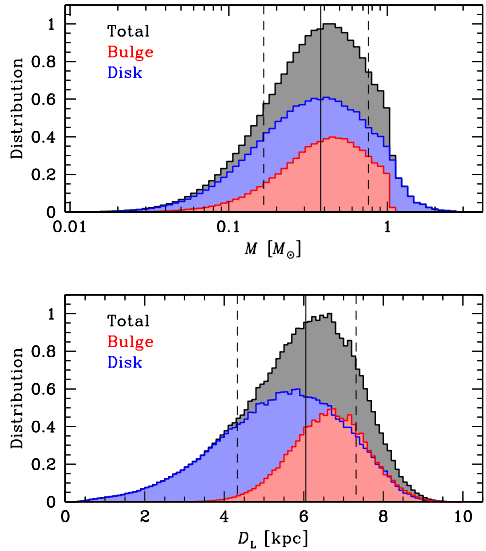}
\caption{Bayesian estimates, based on the \citet{han95} Galactic model,
for the host mass and system distance of KMT-2016-BLG-0212 for
the ``wide 2a'' (sub-Neptune-class) solution.  The distributions are
qualitatively similar to those of the ``close 1'' solution 
(Figure~\ref{fig:bayesc1}).  They are also extremely similar to the
distributions for the ``wide 2b'' and ``wide 3'' solutions, for which
reason these latter are not shown.
\label{fig:bayesw2a}}
\end{figure}

We can nevertheless make a Bayesian estimate of the physical parameters,
i.e., the lens mass $M$, the companion mass $m_p$, the lens distance $D_L$
and the companion-host projected separation $a_\perp$.  To do
so, we draw lens and source kinematics randomly from 
a \citet{han95} Galactic model and draw host masses randomly from
a \citet{chabrier03} mass function.  We then weight these by how well they match the
measured $t_\e$ and $\mu$ and also by the microlensing rate 
$\propto \mu\theta_\e$.  The distributions of the host-mass and system
distance for the two classes of solutions are illustrated in 
Figures~\ref{fig:bayesc1} and \ref{fig:bayesw2a}.  These distributions
peak near $M\sim 0.5\,M_\odot$ and $D_L\sim 6\,\kpc$, but are quite
broad.
[h]

[h]

We find estimated parameter ranges as described in Table~\ref{tab:quant}.  
As expected, the companion of ``close 1'' (``BD'') solution peaks at a
value typical of low-mass BDs, although it overlaps the ``traditional
planetary'' range $m_p<13\,M_{\rm jup}$.  The companions for the three
``Class II'' (``sub-Neptune'') solutions peak in the Super-Earth
regime but are also quite broad.  In Section~\ref{sec:resolve},
we discuss how these two classes of solutions can be distinguished
by future high-resolution imaging of the source.  
While this is the main outstanding
issue in the interpretation of this event, we note that by also resolving
the lens, such observations would simultaneously allow much more
precise determination of the host mass than is returned from the
Bayesian analysis presented in this section.  For the case that
such imaging favors the ``Class I'' solution, the mass of the companion
will be determined quite precisely.  However, for the ``Class II'' solutions,
the companion mass will still be uncertain by a factor $\sim 2$ because
the values of $q$ differ by this amount between solution ``wide 2b''
on the one hand, and the solutions ``wide 2a'' and ``wide 3'' on the other.
See Table~\ref{tab:wide}.

\section{{Future Resolution}
\label{sec:resolve}}

The $q=0.037$ (``brown dwarf'') solution is favored over the
$q\lesssim 10^{-4}$ (``sub-Neptune'') solutions by $\Delta\chi^2=6.64$,
which formally corresponds to $p=\exp(-\Delta\chi^2/2)= 3.6\%$.  This would
not be enough to decisively rule out the latter even if the
statistics of the data were strictly Gaussian.
Moreover,
systematics at this level are quite common in microlensing.  However,
because these two classes of solutions differ in their source flux
by $\Delta I_s \simeq 1\,$mag, it will be straightforward to distinguish
between them with high-resolution imaging, i.e., either adaptive optics (AO)
imaging from the ground or with a high-resolution space telescope,
e.g., the {\it Hubble Space Telescope (HST)}.  This can certainly be
done once the source and lens separate, but there is a good chance
that an observation taken immediately could distinguish between the
two classes of solutions.  In particular, if the flux at the position
of the source 
is significantly below the level expected for the brighter (``sub-Neptune'')
solution, then this would confirm the fainter (``brown dwarf'') solution.
However, if the measured flux is consistent with or brighter than
the brighter solution, then the excess may be due to the lens (or possibly
a companion to the source or the lens).  In this case, additional observations
would be required once the source and lens have separated.

One potential difficulty with ground-based AO observations is that these
are essentially always done in the near-IR.  Since we do not have
a good measurement of the source color, we cannot directly predict
the source flux in IR bands.  However, if the AO observations were 
conducted in two IR band passes (e.g., $J$ and $K$), then the $I$-band
source flux could be determined by making use of an $IJK$ color-color
diagram.  For reference, we note that based on OGLE-III data 
\citep{oiiicat} within $75^{\prime\prime}$ of the lens, the clump
lies at $[(V-I),I]_{\rm clump} = (2.52,16.06)$.

\acknowledgments

Work by WZ, YKJ, and AG were supported by AST-1516842 from the US NSF.
WZ, IGS, and AG were supported by JPL grant 1500811.  
Work by C.H. was supported by the grant (2017R1A4A1015178) of
National Research Foundation of Korea.
This research has made use of the KMTNet system operated by the Korea
Astronomy and Space Science Institute (KASI) and the data were obtained at
three host sites of CTIO in Chile, SAAO in South Africa, and SSO in
Australia.

\end{document}

%% file: tab_c.tex
\begin{table*}
\caption{\textsc{Lensing parameters of close models}} 
\begin{center}
\begin{tabular}{lcccc}
\hline
\hline
\multicolumn{1}{c}{Parameters} & 
\multicolumn{1}{c}{close 1} & 
\multicolumn{1}{c}{close 2} & 
\multicolumn{1}{c}{close 3} & 
\multicolumn{1}{c}{close 4} \\
\hline
$\chi^2$                    & 9150.54            & 9216.07            & 9173.89            & 9316.97            \\
$\rm{dof}$                  & 9147               & 9147               & 9147               & 9147               \\
$t_0$ $(\rm{HJD}^{\prime})$ & 7463.052$\pm$0.249 & 7465.385$\pm$0.193 & 7465.209$\pm$0.195 & 7463.171$\pm$0.260 \\
$u_0$                       & 0.328$\pm$0.013    & 0.591$\pm$0.037    & 0.515$\pm$0.029    & 0.418$\pm$0.020    \\
$t_{\rm E}$ $(\rm{days})$   & 26.616$\pm$0.880   & 17.860$\pm$0.763   & 19.896$\pm$0.781   & 19.906$\pm$0.718   \\
$s$                         & 0.829$\pm$0.007    & 0.709$\pm$0.013    & 0.735$\pm$0.011    & 0.791$\pm$0.005    \\
$q$ $(10^{-4})$             &   368$\pm$61       & 8.889$\pm$1.395    & 3.054$\pm$0.475    &  2467$\pm$205      \\
$\alpha$ $(\rm{rad})$       & 3.131$\pm$0.060    & 4.044$\pm$0.017    & 4.168$\pm$0.017    & 5.086$\pm$0.034    \\
$\rho$ $(10^{-3})$          & 1.192$\pm$0.207    & 1.173$\pm$0.203    & 1.371$\pm$0.218    & --                 \\
\hline
\label{tab:close}
\end{tabular}
\end{center}
\end{table*}

%% file: tab_w.tex
\begin{table*}
\caption{\textsc{Lensing parameters of wide models}}
\begin{center}
\begin{tabular}{lccccc}
\hline
\hline
\multicolumn{1}{c}{Parameters} & 
\multicolumn{1}{c}{wide 1} & 
\multicolumn{1}{c}{wide 2a} & 
\multicolumn{1}{c}{wide 2b} & 
\multicolumn{1}{c}{wide 3} & 
\multicolumn{1}{c}{wide 4} \\
\hline
$\chi^2$                    & 9183.84            & 9157.18            & 9159.21            & 9158.52            & 9214.34            \\
$\rm{dof}$                  & 9147               & 9147               & 9147               & 9147               & 9147               \\
$t_0$ $(\rm{HJD}^{\prime})$ & 7466.971$\pm$0.217 & 7465.316$\pm$0.192 & 7465.250$\pm$0.197 & 7465.276$\pm$0.189 & 7466.843$\pm$0.204 \\
$u_0$                       & 0.224$\pm$0.010    & 0.615$\pm$0.020    & 0.617$\pm$0.022    & 0.619$\pm$0.018    & 0.160$\pm$0.014    \\
$t_{\rm E}$ $(\rm{days})$   & 36.288$\pm$1.210   & 17.764$\pm$0.476   & 17.801$\pm$0.478   & 17.584$\pm$0.440   & 38.853$\pm$2.457   \\
$s$                         & 1.070$\pm$0.005    & 1.427$\pm$0.014    & 1.430$\pm$0.015    & 1.434$\pm$0.012    & 1.685$\pm$0.060    \\
$q$ $(10^{-4})$             &   359$\pm$39       & 0.490$\pm$0.079    & 0.828$\pm$0.153    & 0.484$\pm$0.110    &  1153$\pm$216      \\
$\alpha$ $(\rm{rad})$       & 0.013$\pm$0.022    & 1.004$\pm$0.0145   & 1.001$\pm$0.0153   & 1.005$\pm$0.014    & 0.867$\pm$0.022    \\
$\rho$ $(10^{-3})$          & 0.791$\pm$0.139    & 1.378$\pm$0.278    & 1.666$\pm$0.283    & 1.874$\pm$0.430    & --                 \\
\hline
\label{tab:wide}
\end{tabular}
\end{center}
\end{table*}

%% file: tab_quant.tex
\begin{table*}
\caption{\textsc{Physical properties}} 
\begin{center}
\begin{tabular}{lcccc}
\hline
\hline
\multicolumn{1}{c}{Quantity} & 
\multicolumn{1}{c}{close 1} & 
\multicolumn{1}{c}{wide 2a} & 
\multicolumn{1}{c}{wide 2b} & 
\multicolumn{1}{c}{wide 3} \\
\hline
I$_{\rm s}$ - I$_{\rm clump}$  & 4.645$\pm$0.080  & 3.558$\pm$0.077  & 3.551$\pm$0.077   & 3.550$\pm$0.077   \\
(V-I)$_{\rm s}$ - (V-I)$_{\rm clump}$     & -0.29$\pm$0.12      & -0.23$\pm$0.11   & -0.23$\pm$0.11  & -0.24$\pm$0.11  \\
$\theta_{\rm E}$ [mas] &  0.42$\pm$0.10      &  0.64$\pm$0.22   &  0.53$\pm$0.14     &  0.47$\pm$0.15    \\
$\mu$ [mas]          &   8.1$\pm$1.9       &  13.2$\pm$4.5    &  10.9$\pm$2.9      &   9.8$\pm$3.1     \\
$M$ [$M_\odot$]       &  0.49$_{-0.26}^{+0.37}$ & 0.38$_{-0.22}^{+0.38}$ &  0.41$_{-0.23}^{+0.38}$  & 0.373$_{-0.21}^{+0.37}$ \\
$m_{\rm p}$ & 19$_{-12}^{+20}\,M_{\rm Jup}$ & 6.2$_{-4.0}^{+8.2}\,M_\oplus$ & 11.3$_{-7.2}^{+14.4}\,M_\oplus$ & 6.0$_{-4.0}^{+8.6}\,M_\oplus$ \\
$D_{\rm L}$ [kpc]      &  6.3$_{-1.5}^{+1.1}$   & 6.1$_{-1.7}^{+1.3}$ &  5.9$_{-1.6}^{+1.3}$  & 6.2$_{-1.6}^{+1.2}$ \\
$a_\perp$ [AU]         &  2.2$_{-0.5}^{+0.5}$   & 5.5$_{-1.9}^{+2.0}$ &  4.5$_{-1.2}^{+1.2}$  & 4.2$_{-1.4}^{+1.4}$ \\
\hline
\label{tab:quant}
\end{tabular}
\end{center}
\end{table*}